\documentclass{WileyMSP-template}

\usepackage{graphicx}
\usepackage{dcolumn}
\usepackage{bm}
\usepackage[colorlinks=true,citecolor=blue,linkcolor=magenta]{hyperref}
\usepackage[utf8]{inputenc}
\usepackage{url}
\usepackage[normalem]{ulem}
\usepackage{upgreek}
\usepackage[mathlines]{lineno}
\usepackage{amsmath}
\usepackage{lineno}
\usepackage{ragged2e} 

\begin{document}

\pagestyle{fancy}
\rhead{\includegraphics[width=2.5cm]{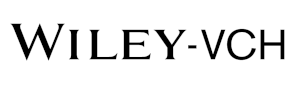}}

\title{Chip-based $f$--2$f$ interferometry in periodically tapered lithium niobate nanophotonic waveguides}

\maketitle


\author{Xinyan Chi$^{1}$,}
\author{Ruoao Yang$^{2}$,}
\author{Zhiyuan Li$^{1}$,}
\author{Tuo Liu$^{1}$,}
\author{Haoxuan Zhang$^{1},$}
\author{Biyan Zhan$^{1}$,}
\author{Xianwen Liu$^{1,3,*}$}


\begin{affiliations}
\normalsize
1 Beijing Engineering Research Center of Mixed Reality and Advanced Display, MIIT Key Laboratory of Photonics Information Technology, School of Optics and Photonics, Beijing Institute of Technology, Beijing, 100081, China\\
2 State Key Laboratory of Nuclear Physics and Technology, Key Laboratory of HEDP of the Ministry of Education, CAPT, School of Physics, Peking University, Beijing 100871, China\\
3 National Key Laboratory on Near-surface Detection, 100072, Beijing, China\\
$\star$ Email Address: xianwen.liu@bit.edu.cn
\end{affiliations}



\begin{abstract}
\justifying
Nanophotonic supercontinuum generation offers a practical route to chip-based $f$--2$f$  interferometry by leveraging coexisting $\chi^{(2)}$ and $\chi^{(3)}$ nonlinearities. In conventional uniform waveguides, the phase-matching bandwidth for second-harmonic generation (SHG) is intrinsically narrow, restricting the spectral overlap factor for heterodyne beating. To address this limitation, we introduce a periodically tapered nanophotonic waveguide made from MgO-doped, z-cut thin-film lithium niobate for energy-efficient and fabrication-robust $f$--2$f$ operation. By adiabatically varying the waveguide width within a dual phase-matching window that supports concurrent dispersive wave (DW) emission and SHG, we routinely achieved a broad spectral overlap between the SHG and DW components. This capability enables robust detection of the carrier–envelope offset frequency ($f_\mathrm{ceo}$) at substantially lower pulse energies than that in uniform-waveguide approaches. We further developed a compact waveguide module that operates reliably under temperature fluctuations and is capable of interfacing with high-repetition-rate (500\,MHz) mode-locked lasers, enabling detection and phase locking of $f_\mathrm{ceo}$ with a signal-to-noise ratio of 48\,dB. These results highlight the potential of nanophotonic chips for developing compact, field-deployable frequency comb systems.
\end{abstract}

\section{Introduction}
\justifying
\hspace{2em}Optical frequency combs are indispensable tools for precision metrology and timekeeping, where absolute frequency traceability are paramount \cite{Fortier2019,Diddams2020}. Practical deployment requires full comb stabilization, which in turn demands reliable extraction of its carrier–envelope offset frequency ($f_\mathrm{ceo}$) \cite{jones2000carrier}. This is commonly achieved via a $f$--2$f$ interferometer, where the long-wavelength portion of an octave-spanning comb is frequency-doubled and heterodyned with its short-wavelength counterpart \cite{Telle1999,Tian2024}. Achieving such octave broadening traditionally relies on supercontinuum generation (SCG) in highly nonlinear fibers \cite{dudley2006supercontinuum}. Nanophotonic waveguides offer a compact, energy-efficient alternative owing to their subwavelength optical confinement and lithographically programmable dispersion \cite{yang2024generation,zia2023ultraefficient,lee2024inverse}, allowing ultra-broadband SCG on millimeter-scale chips with pulse energies of a few hundred picojoules or less \cite{bres2023supercontinuum,gaeta2019photonic,guo2018mid,lu2020ultraviolet}.

Moreover, chip-scale $f$--2$f$  interferometers are feasible in photonic platforms that concurrently host $\chi^{(2)}$ and $\chi^{(3)}$ nonlinearities \cite{liu2023aluminum,zhu2021integrated}, allowing SCG and second-harmonic generation (SHG) to be co-implemented on a single chip, hence eliminating discrete frequency-doubling components \cite{Mayer2015,lamee2020nanophotonic}. This approach has been demonstrated across several material platforms, including aluminum nitride (AlN) \cite{hickstein2017ultrabroadband}, gallium nitride (GaN) \cite{fan2024supercontinua}, silicon nitride (SiN) \cite{Okawachi2018,Hickstein2019}, and thin-film lithium niobate (TFLN) \cite{Yu2019,okawachi2020chip,obrzud2021stable}. Among these, TFLN is particularly compelling because it offers strong $\chi^{(2)}$ nonlinearities ($d_{33}$\,=\,19.5\,pm/V, $d_{31}$\,=\,3.2\,pm/V@1.3\,$\mu$m), flexible ferroelectric domain-engineering capabilities, and ready access to high-quality commercial wafers \cite{Shoji1997,lu2019periodically,Jia2021}. Although cascaded $\chi^{(2)}$ interactions have enabled on-chip $f_\mathrm{ceo}$ detection at picojoule to sub-hundred-picojoule pulse energies \cite{Jankowski2020,Hamrouni2024,wu2024visible,ayhan2025fabrication}, the fabrication of high-quality periodically poled TFLN waveguides is nontrivial, requiring tight control over waveguide parameters and poling periods \cite{chen2024adapted,li2024advancing}. An appealing alternative is SHG via mode phase-matching (MPM) without resorting to the  poling process. The strategy is to tailor the waveguide dimensions such that $\chi^{(3)}$-mediated SCG  produces dispersive waves (DWs) at a wavelength that overlaps spectrally with $\chi^{(2)}$-driven SHs \cite{okawachi2020chip}. Owing to millimeter-scale interaction lengths, the group-velocity mismatch (or temporal walk-off) between DW and SH fields can be sufficiently small or negligible. Therefore, efficient $f$--2$f$ interferometry becomes feasible. 

In uniform waveguides, MPM-based SHG typically suffers from a narrow phase-matching bandwidth \cite{luo2018highly}, which only spectrally overlaps with part of the broadband DW produced during SCG \cite{Yu2019}. This bandwidth mismatch reduces the available heterodyne power, therefore degrading the $f_\mathrm{ceo}$ beat-note efficiency, especially at low pulse energies as preferred for cost-efficient frequency comb systems. In addition, the SHG phase-matching wavelength is susceptible to nanometer-scale fabrication deviations (e.g., waveguide width, film thickness, and etching depth) \cite{wang2017second}, imposing stringent demands on process control in uniform waveguides. To mitigate this issue, our earlier work proposed a chirp-modulated AlN tapered waveguide that supports broad, gap-free SHG extending into the ultraviolet \cite{liu2019beyond}, suggesting the feasibility of achieving broadband $\chi^{(2)}$ frequency conversion in poling-free nanophotonic waveguides.

\begin{figure}[!t]
	\centering
	\includegraphics[width=0.5\linewidth]{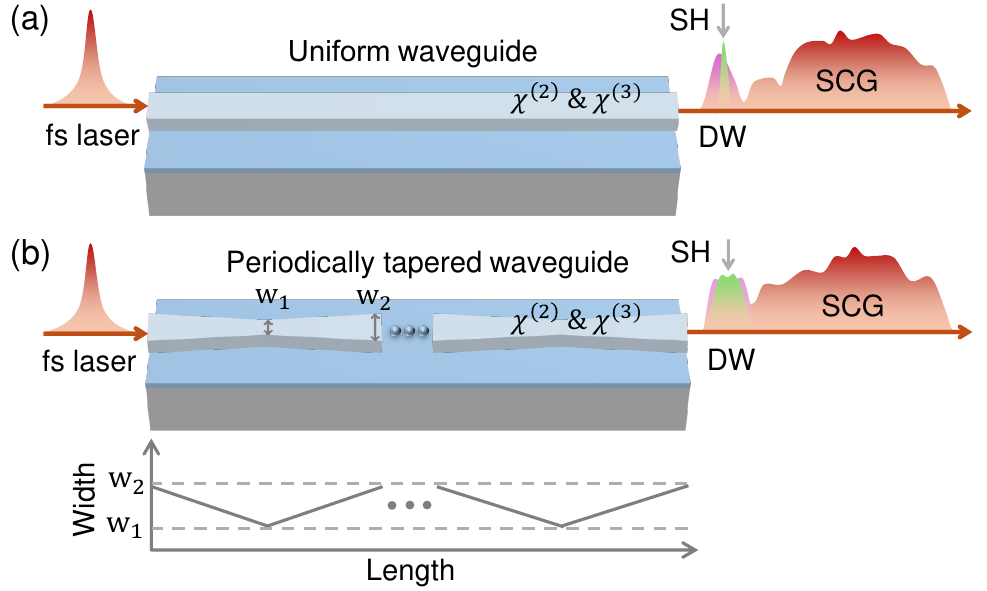}
	\caption{(a, b) Schematic of spectral broadening via $\chi^{(2)} $ and $\chi^{(3)} $ nonlinearities in femtosecond (fs) laser-driven uniform and periodically tapered nanophotonic waveguides, respectively. The degree of spectral overlap between the SH and DW components is significantly enhanced in (b). Bottom of (b): sketch of the varied waveguide width along the propagation direction.} 
	\label{Fig1}
\end{figure}

In this work, we demonstrate a novel periodically tapered nanophotonic waveguide that enables chip-based $f$--$2f$ interferometry with reduced energy requirement and enhanced fabrication tolerance. The devices were designed and fabricated in z-cut, 5mol.\% MgO-doped TFLN (MgO:LN hereafter). By tailoring the phase-matching condition, we achieved a broad spectral overlap between the SH and DW components, enabling robust retrieval of the $f_\mathrm{ceo}$ beatnote at substantially lower pulse energies than that in uniform-waveguide approaches. We further developed a compact, fiber-coupled waveguide module for plug-and-play operation, which remains reliable under temperature fluctuations from 20 to 35$^\circ$C, as limited by our measurement. The module is also capable of interfacing with a high-repetition-rate (500\,MHz) femtosecond laser, delivering an $f_\mathrm{ceo}$ beatnote with a signal-to-noise ratio (SNR) of 48\,dB, which was phase-locked using a servo feedback loop. 

\section{Device design and fabrication}
\hspace{2em}Figure\,\ref{Fig1}(a) sketches a uniform waveguide that supports both $\chi^{(2)}$ and $\chi^{(3)}$ nonlinearities for chip-scale $f$--$2f$ interferometry. In this case, narrow-band SHs spectrally coincide with broadband DWs when their respective phase-matching conditions are satisfied. Because a large portion of DWs remains outside the overlap region, the optical power contributing to the $f_\mathrm{ceo}$ beatnote is intrinsically limited. In addition, realistic fluctuations in film thickness, etch depth, and waveguide width can measurably shift the SHG wavelength, degrading the device yield. To overcome these constraints, we propose a periodically tapered waveguide shown in Fig.\ref{Fig1}(b), where the width is linearly varied along the propagation direction with a predefined period and modulation depth. This design enables adiabatic phase matching for broadband SHG, thereby increasing the spectral overlap with DWs while simultaneously improving tolerance to realistic dimensional offsets. Consequently, efficient and robust $f$--$2f$ interferometers can be realized.

\begin{figure*}[!t]
	\centering
	\includegraphics[width=0.85\linewidth]{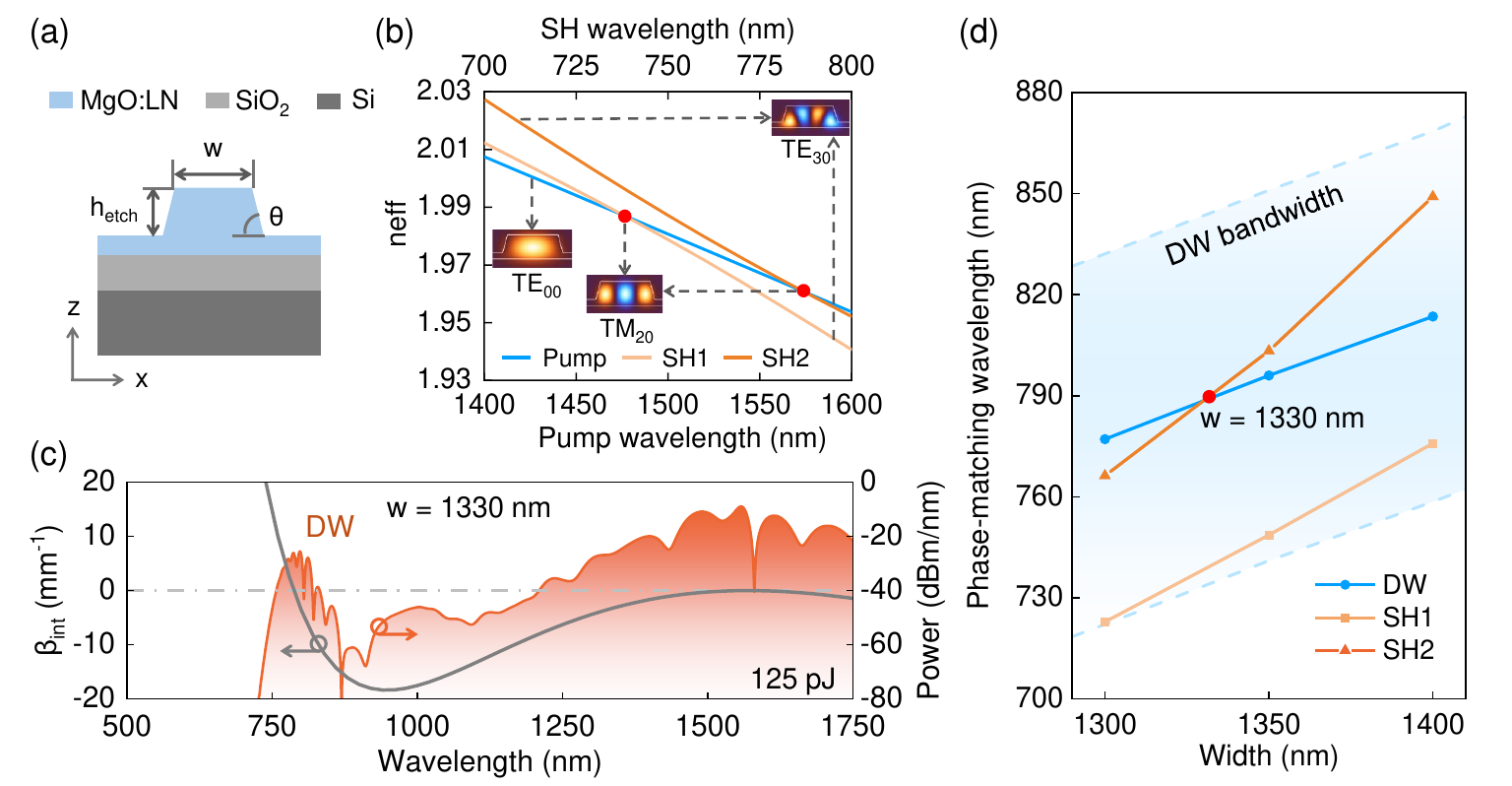}
	\caption{(a) Cross-sectional dimensions of the z-cut MgO:LN waveguide. (b) Effective refractive indices ($n_\mathrm{eff}$) of TE$_{00}$ pump (blue solid) and TM$_{20}$ SH (orange solid) modes in the MgO:LN waveguide (w\,=\,1330 nm). Inset: their corresponding electric field profiles. (c) Simulated integrated dispersion ($ \beta\mathrm{_{int}}$) and the corresponding supercontinuum spectrum in the MgO:LN waveguide at an on-chip pulse energy of 125\,pJ. (d) Phase-matching wavelengths of DW emission and SHG versus the waveguide width with other optimal parameters being fixed. The shaded window indicates a 20-dB DW spectral bandwidth.} 
	\label{Fig2}
\end{figure*}

We designed the waveguide on a 600-nm-thick, z-cut MgO:LN film atop a 2-$\mu$m silicon dioxide layer and a 525-$\mu$m silicon handle. Compared with undoped TFLN, MgO doping significantly increases the optical-damage threshold \cite{nakamura2002optical}, enabling stable, long-term operation under pulsed excitation. For chip-scale $f$--$2f$ interferometry, we tailored the cross-section of MgO:LN waveguide [Fig.\,\ref{Fig2}(a)]  to satisfy a dual phase-matching condition that enforces spectral overlap between the DW and SH. Through finite-element simulations, we identified an optimal waveguide geometry with a top width of 1330\,nm, an etch depth of 450\,nm, and a sidewall angle of 75$^{\circ}$. As shown in Fig.\,\ref{Fig2}(b), this design supports MPM-based SHG via the $d_{31}$ tensor element, where the effective indices ($n_\mathrm{eff}$) of the fundamental pump mode (TE$_{00}$) and the second-order SH mode (TM$_{20}$) coincide at specific wavelengths. Owing to the intrinsic birefringence of MgO:LN, mode hybridization between the TM$_{20}$ and TE$_{30}$ modes arises in the region of  700--800\,nm, inducing two distinct SH branches (SH1 and SH2), whose $n_\mathrm{eff}$ curves intersect that of the pump. Consequently, two SH peaks near 738\,nm and 787\,nm can be expected, as verified by our later experimental results.

For soliton-induced DW emission, the phase mismatch between the DW and the pump pulse can be quantified by the integrated dispersion ($\beta_\mathrm{int}$) \cite{guo2018mid}:
\begin{equation}
	\beta_\mathrm{{int}} = \beta\left(\omega\right)- \beta\left( \omega_{0}\right)-\left(\omega-\omega_{0} \right)\nu\mathrm{_{g}}^{-1}
	\label{eq.1}
\end{equation}
Here, $\beta(\omega) $ denotes the propagation constant, $ \omega_{0} $ is the pump angular frequency, and $\nu\mathrm{_{g}}$ is the group velocity at $\omega_{0}$. The phase-matched DW wavelength can be numerically predicted from the zero-crossing condition ($\beta\mathrm{_{int}}$\,=\,0). As shown in Fig.\,\ref{Fig2}(c), for the optimized waveguide geometry, the calculated $\beta\mathrm{_{int}}$ predicts a DW near 787\,nm (i.e., $\beta_\mathrm{int}$\,=\,0), which is consistent with SCG simulation using a generalized nonlinear Schrödinger equation \cite{agrawal2011nonlinear} (see Supplementary Note I for details). Notably, the predicted DW wavelength coincides with the SHG phase-matching point based on our design.

In practical nanofabrication, achieving tight, repeatable control of the waveguide width, etch depth, and film thickness remains nontrivial. As presented in Fig.\,\ref{Fig2}(d), even modest width deviations produce measurable shifts in the phase-matched wavelengths for both SHG and DW emission. Fortunately, once the pulse energy is sufficient to drive SCG, the DW exhibits a broad spectral envelope [see Fig.,\ref{Fig2}(c)] that can retain overlap with the narrow SH, thereby relaxing the demand for exact structural fidelity.  In addition, variations in etch depth and film thickness can be mitigated by lithographically retuning the waveguide width. As detailed in Supplementary Note II, an appropriate width adjustment can recover a common phase-matched wavelength for the DW and SH across etch depths of 450--500\,nm and film thicknesses of 590--610\,nm. These factors enable chip-based $f$--$2f$ interferometry in uniform nanophotonic waveguides with both $\chi^{(2)}$ and $\chi^{(3)}$ nonlinearities \cite{hickstein2017ultrabroadband,fan2024supercontinua,Okawachi2018,Hickstein2019,Yu2019,okawachi2020chip,obrzud2021stable}.

Guided by this approach, we fabricated a set of MgO:LN uniform waveguides with lithographically varied widths to alleviate process-induced parameter offsets. To enable robust operation, we adopted a periodically tapered geometry, in which the patterned waveguide width (w$_\mathrm{p}$) is linearly swept from 1330\,nm to 1440\,nm and repeated with a 1\,mm period along the propagation direction. The selected sweep range is motivated by the broad DW bandwidth in Fig.\,\ref{Fig2}(d), and we include  an additional 40-nm width offset to better align the design with our realistic fabrication. In contrast to a single taper structure, the cyclic width modulation reduces sensitivity to the film thickness nonuniformity by repeatedly re-establishing local phase-matching conditions. 

The nano-scale waveguide patterns were defined by 100-kV electron-beam lithography using positive-tone ZEP520A resists and transferred into the MgO:LN layer using angled Ar$^{+}$ milling, yielding a sidewall angle of $\sim$75$^\circ$ \cite{zhan2025inverse}. Following dry etching, the chip was cleaned in a standard SC1 solution (NH$_4$OH:H$_2$O$_2$:H$_2$O
=1:1:5) at 60\,$^\circ$C for 10 minutes to remove resist residues and etch by-products. Finally, the chip was annealed at 500\,$^\circ$C for three hours to mitigate fabrication-induced defects and then cleaved to expose the waveguide facets, yielding a total length of 7\,mm. To improve end-fire coupling efficiency, both the uniform and periodically tapered waveguides incorporate 3-$\mu$m-wide access sections at the input and output facets. 

\section{Results and Discussion}
\begin{figure*}[!h]
	\centering
	\includegraphics[width=1\linewidth]{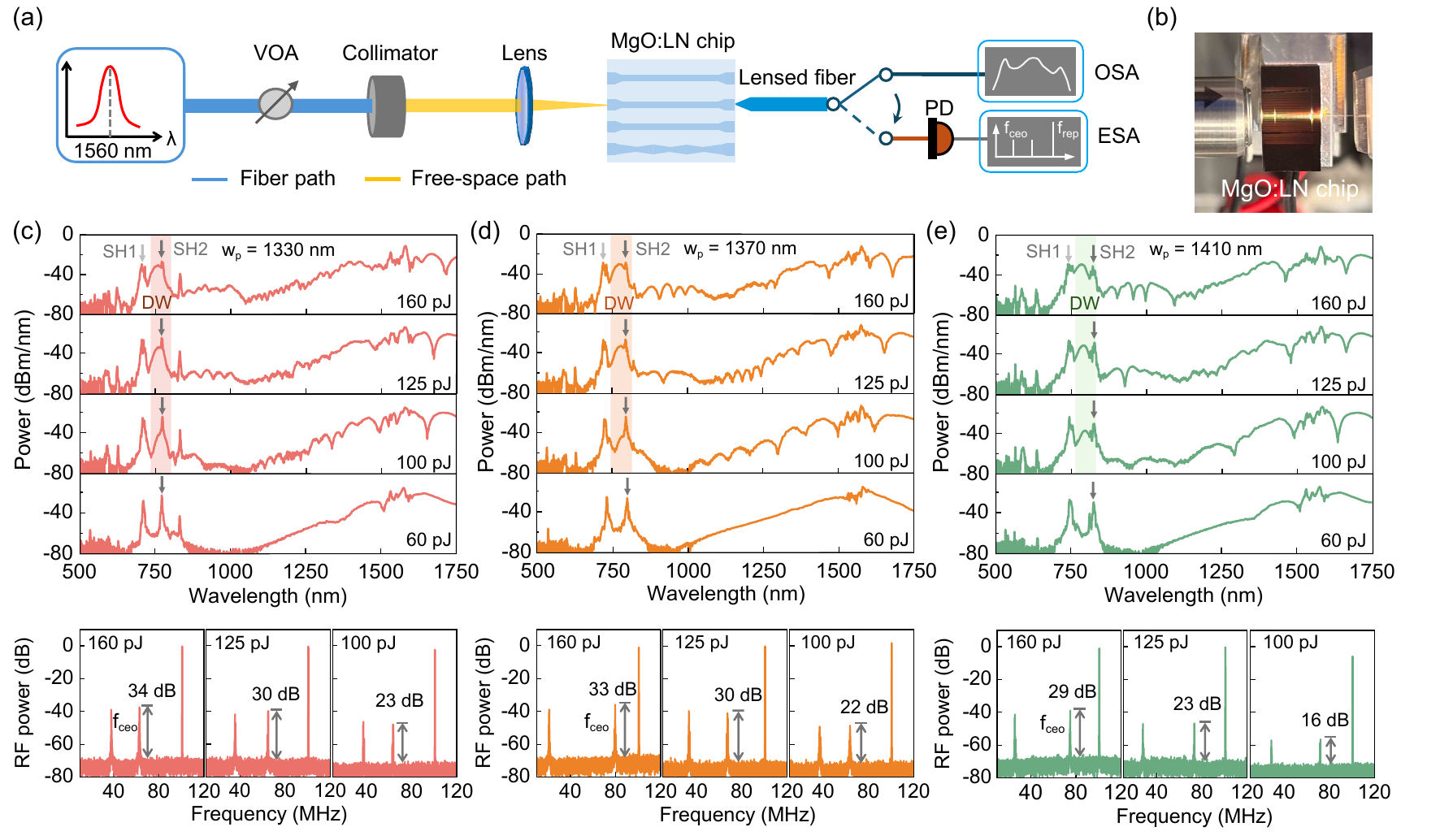}
	\caption{(a) Schematic of the experimental setup. (b) Photograph of the MgO:LN chip under test,  showing pronounced visible light scattering. (c--e) Measured optical (top) and the corresponding RF spectra (bottom) of uniform MgO:LN waveguides with patterned  widths (w$_\mathrm{p}$) of 1330, 1370, and 1410\,nm, respectively. The applied on-chip pulse energy varies from 60 to 160\,pJ, while the adopted resolution bandwidth (RBW) and video bandwidth (VBW) are 100\,kHz and 10\,kHz, respectively. The repetition-rate beatnote is near 100\,MHz.}
	\label{Fig3}
\end{figure*}

\hspace{2em}Figure\,\ref{Fig3}(a) illustrates our experimental setup for evaluating the device performance. The pump source was a compact erbium-doped fiber femtosecond laser centered near 1560\,nm, delivering $\sim$100\,fs pulses at a repetition rate of $\sim$100\,MHz, with up to 120\,mW average power. A fiber-pigtailed variable optical attenuator (VOA) was inserted to regulate the launched power, thereby avoiding the pulse-width variations associated with directly adjusting the laser. 
The attenuated beam was sent to a fiber collimator and focused onto the chip facet using an aspheric lens. Because the input fiber chain was polarization‑maintaining, the pump polarization was set to horizontal for exciting the TE$_{00}$ waveguide mode by rotating the fiber collimator. The transmitted light was collected with a single-mode lensed fiber and routed to a grating-based optical spectrum analyzer (OSA, 350--1750\,nm) for spectral characterization. In parallel, the $f_\mathrm{ceo}$ beatnote was acquired by directing the output to a silicon photodetector (PD, LBTEK PDB200A-DC, 200\,MHz) following by an electrical spectrum analyzer (ESA). Unlike earlier implementations using free-space optical filters \cite{fan2024supercontinua,Okawachi2018,Hickstein2019,Yu2019,okawachi2020chip}, we employed a 780\,nm single-mode fiber to suppress the spectrum above 1000\,nm.

Figure\,\ref{Fig3}(b) presents a photograph of the MgO:LN chip under test. When pumped by a femtosecond laser near 1560\,nm, we observed bright visible light scattering from the chip surface, providing direct evidence of DW emission and SHG in the waveguide. The insertion losses were calibrated to be $\sim$5.0\,dB and 4.5\,dB for free-space and lensed fiber coupling, respectively. By varying the on-chip pulse energy, we have recorded the broadened optical spectra and the corresponding RF beatnotes for uniform waveguides with patterned widths of w$_\mathrm{p}$\,=\,1330, 1370, and 1410\,nm, respectively. The results are shown in Figs.\,\ref{Fig3}(c--e). At 60\,pJ, the optical spectra exhibit self-phase-modulation (SPM) broadening around the pump, accompanied by two pronounced visible SH peaks (SH1 and SH2), in agreement with the phase-matching design in Fig.\,\ref{Fig2}(b). The $f_\mathrm{ceo}$ beatnote was not observed due to insufficient pump energy for enabling DW emission. 

At an on-chip pulse energy of 100\,pJ, soliton fission is initiated and DW emerges with a modest optical power, resulting in the captured $f_\mathrm{ceo}$ beat-note SNRs below 23\,dB. The DW formation was verified through numerical simulations with varied pulse energies (see Supplementary Note I). As the pulse energy is increased to 125\,pJ and 160\,pJ,  both the DW power and spectral bandwidth enhance, yielding substantially improved overlap with the SH peaks. Under these conditions, the obtained $f_\mathrm{ceo}$ beat-note SNR reaches 34\,dB for w$_\mathrm{p}$\,=\,1330 and 1370\,nm, while remaining below 30\,dB for w$_\mathrm{p}$\,=\,1410\,nm. This geometry-dependent behavior arises from the spectral detuning between the DW and SH, which red-shift at different rates with increasing waveguide widths, consistent with the calculated phase-matching trend in Fig.\,\ref{Fig2}(d).
\begin{figure*}[!h]
	\centering
	\includegraphics[width=0.85\linewidth]{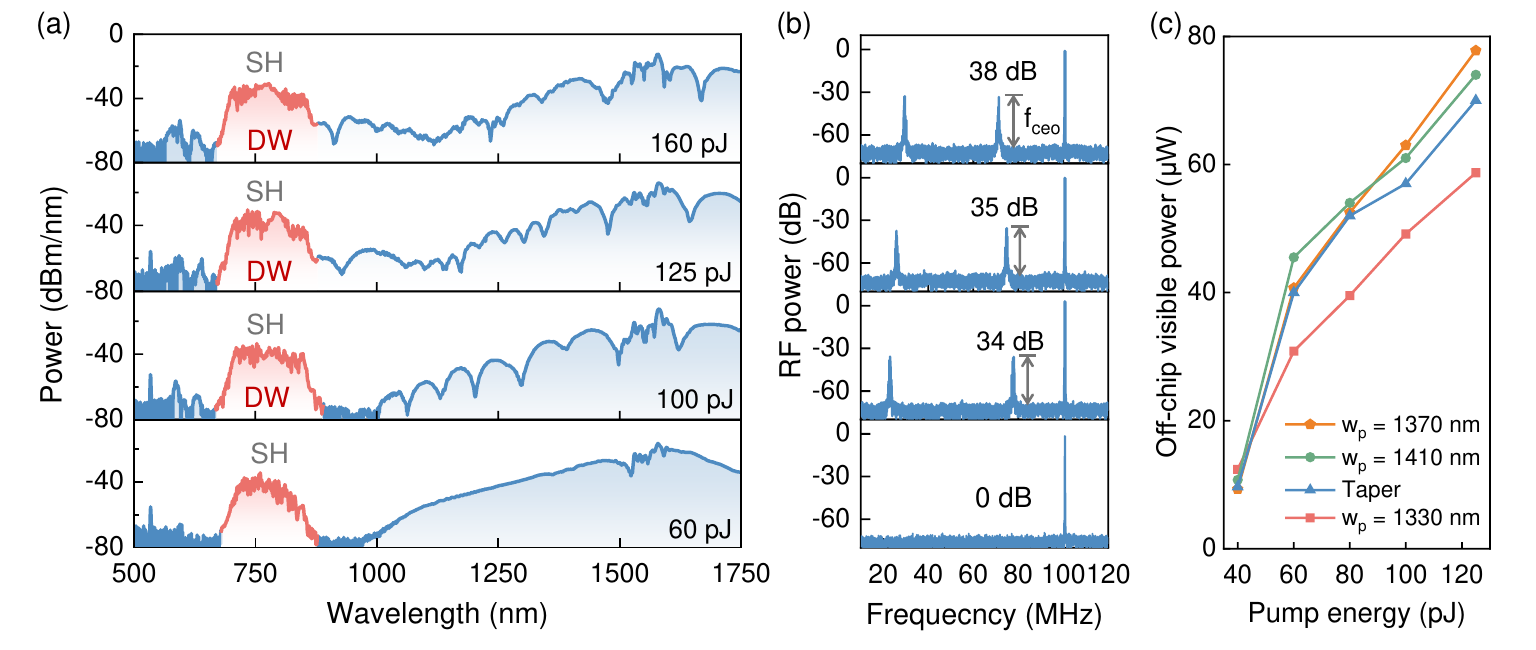}
	\caption{(a, b) Evolution of the optical spectrum and the corresponding RF beatenotes (RBW\,=\,100 kHz, VBW\,=\,10 kHz) from the periodically tapered MgO:LN waveguide. The applied on-chip pulse energy varies from 60 to 160\,pJ, and the repetition-rate beatnote is near 100\,MHz. (c) Comparison of off-chip visible power from the uniform and periodically tapered waveguides versus the on-chip pump energies.}
	\label{Fig4}
\end{figure*}

Figure\,\ref{Fig4}(a) presents the optical spectra from a periodically tapered MgO:LN waveguide that was engineered to realize adiabatic phase matching for broadband SHG. At a modest pulse energy of 60\,pJ, we obtained an SH bandwidth exceeding 200\,nm, an order-of-magnitude broader than that in co-fabricated uniform waveguides. The broadband SHG response was corroborated by complementary continuous-wave measurements (see Supplementary Note III). Since the pulse energy of 60\,pJ is below the threshold for DW emission, no $f_\mathrm{ceo}$ beatnote was observed, as shown in Fig.\,\ref{Fig4}(b). Increasing the pulse energy to 100\,pJ triggers broadband DW generation that spectrally overlaps with the braodband SH, yielding an $f_\mathrm{ceo}$ beat-note SNR of 34\,dB. In contrast, uniform waveguides require a higher pulse energy of 160\,pJ to achieve comparable performance [Fig.\,\ref{Fig3}(c)]. Further increasing the pulse energy in the tapered device produces only a modest SNR improvement, reaching a maximum of 38\,dB, which indicates that the SH-DW spectral overlap is already near saturation.

Beyond improved energy efficiency, the periodically tapered structure is inherently less sensitive to variations in etch depth and film thickness. With this design approach, we reproducibly obtained $f_\mathrm{ceo}$ beatnotes with SNRs exceeding 35\,dB across all fabricated waveguides. Notably, the measured $f_\mathrm{ceo}$ SNR is ultimately limited by the noise performance of our femtosecond pump laser. When a lower-noise source is used, the SNR increases by $\sim$10\,dB (see later discussion). We also compared the visible output power from uniform and periodically tapered MgO:LN waveguides. 
The lensed-fiber output was directed to a silicon optical power meter (Thorlabs S140C, 350--1100nm). As shown in Fig.\ref{Fig4}(c), the periodically tapered device exhibits a comparable visible power with that of the uniform waveguides, especially for on-chip pump energies below 80\,pJ. The result indicate that periodic tapering design can simultaneously preserve the SHG bandwidth and conversion efficiency, consistent with our prior observation \cite{liu2019beyond}.

\begin{figure*}[!h]
	\centering
	\includegraphics[width=1\linewidth]{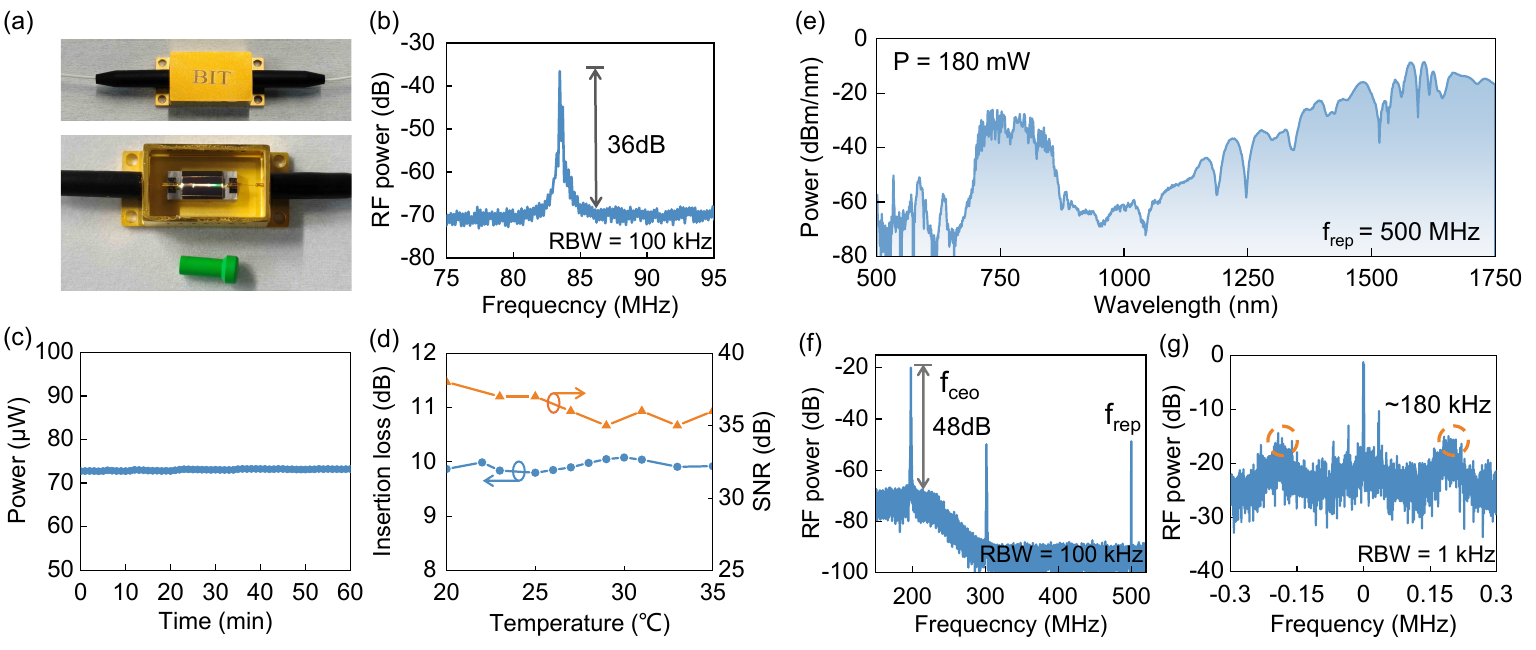}
	\caption{(a) Photograph of the packaged waveguide module with its size compared to a fiber dust cap. Owing to SCG, the chip inside emits a bright visible glow after interfaced by a femtosecond laser. (b) Measured $f_\mathrm{ceo}$ beatnote when feeding the module with a 100\,MHz femtosecond laser at 50\,mW (RBW\,=\,100\,kHz, VBW\,=\,1\,kHz).  (c) Corresponding visible output power recorded over 1 hour. (d) Measured insertion loss and $f_\mathrm{ceo}$ beat-note SNR versus the varied module temperature. (e, f) Optical spectrum and corresponding RF beatnotes under excitation with a 500\,MHz femtosecond laser at 180\,mW. (g) Phase-locked $f_\mathrm{ceo}$ signal (RBW\,=\,1\,kHz, VBW\,=\,100\,Hz) versus the frequency detuning.}
	\label{Fig5}
\end{figure*}

For portable deployment, we developed a compact, fiber-coupled waveguide module (length of 3.5\,cm), as shown in Fig.\ref{Fig5}(a). To improve the coupling efficiency while preserving a horizontal polarization, we interfaced the MgO:LN waveguide using polarization-maintaining lensed fibers with a 2.5\,$\mu$m spot diameter at 1550\,nm. After precise alignment, the lensed fibers equipped with metal ferrules were laser-welded to the submount, forming a mechanically rigid, vibration-tolerant assembly. The packaged module exhibits an insertion loss of 4.9\,dB/facet, only 0.4\,dB higher than the bare-chip measurement. By co-integrating inverted-taper mode-size converters \cite{He2019}, the insertion loss can be  further reduced.

To validate the module performance, we pumped it with the 100\,MHz femtosecond laser described above. We captured an $f_\mathrm{ceo}$ beatnote with an SNR of 36\,dB [Fig.\ref{Fig5}(b)]. The correspond visible output power was recorded in Fig.\,\ref{Fig5}(c), which remains essentially constant over a one-hour interval under typical laboratory conditions. To quantify thermal robustness, we mounted the module on a thermoelectric cooler (TEC) and varied the base temperature from 20\,$^{\circ}$C to 35\,$^{\circ}$C. As shown in Fig.\ref{Fig5}(d), the fiber-to-fiber insertion loss varies by less than 0.3\,dB, while the $f_\mathrm{ceo}$ SNR stays above 35\,dB, suggesting the suitability of the module for operating outside laboratory environments. The weak temperature dependence of $f_\mathrm{ceo}$ is consistent with the near-zero thermo-optic coefficient ($dn_\mathrm{o}/dT$\,$\approx$\,0) of z-cut MgO:LN under TE-polarized pumping \cite{luo2018highly}, which renders a modest SHG wavelength shift of 0.37\,nm/$^{\circ}$C, as detailed in Supplementary Note III.

To investigate the scalability of the packaged module for higher repetition-rate laser applications, we pumped it with a home-built 500\,MHz fiber femtosecond laser (pulse duration of $\sim$110\,fs) \cite{Chen2024}. As shown in Fig.\ref{Fig5}(e), the output spectrum retains broadband DW and SH features, closely mirroring the behavior observed under the 100-MHz excitation [Fig.\ref{Fig4}(a)]. After photodetection, we obtained an $f_\mathrm{ceo}$ beatnote with an SNR reaching 48\,dB [Fig.\ref{Fig5}(f)], approximately 10\,dB higher than that in the 100\,MHz case. This discrepancy is attributed to the distinct noise characteristics of the two pump laser, which directly impact the achievable $f_\mathrm{ceo}$ SNR. We further implemented active stabilization of $f_\mathrm{ceo}$ using a commercial servo controller (Vescent D2-135).  The detected $f_\mathrm{ceo}$ beatnote was amplified to 0\,dBm and compared to a reference signal disciplined by a rubidium atomic clock. The resulting phase/frequency error signal was processed by a proportional–integral–derivative (PID) controller and applied as feedback to the laser drive current. Figure\,\ref{Fig5}(g) presents the stabilized $f_\mathrm{ceo}$ near 167\,MHz, from which a locking bandwidth of approximately 180\,kHz is inferred.

\section{Conclusion}
\hspace{2em}We demonstrate a novel chip-based $f$--$2f$ interferometry architecture that simultaneously reduces the required pulse energy and enhances fabrication tolerance. This is enabled by employing a periodically tapered MgO:LN nanophotonic waveguide with coexisting $\chi^{(2)}$ and $\chi^{(3)}$ nonlinearities. By optimizing the waveguide geometry (film thickness, ridge width, and etch depth), we identified a dual phase-matching window for concurrent SHG and DW emission. Unlike uniform waveguides with narrow SH spectral peaks, the periodically tapered device enforces adiabatic phase matching that produces broadband SHG with a measured spectral bandwidth exceeding 200\,nm, while substantially reducing sensitivity to practical dimensional perturbations. The resulting broad spectral overlap between the SH and DW yields an $f_\mathrm{ceo}$ beat-note SNR of 34\,dB at a reduced on-chip pulse energy of 100\,pJ. For portable and plug-and-play operation, we also developed a compact, fiber-coupled waveguide module that remains reliable under temperature variations from  20 to 35$^\circ$C. When driven by a 500\,MHz repetition-rate femtosecond laser, the module delivers an $f_\mathrm{ceo}$ beatnote with an exceptional SNR of 48\,dB, which was phase-locked using a servo feedback loop. Our approach highlights the great potential of nanophotonic chips for developing compact, field-deployable frequency-comb systems.


\medskip
\textbf{Supporting Information}\par
Supporting Information is available from the Wiley Online Library or from the author.

\section*{Conflict of Interest} 
The authors have filed an invention patent application (CN2026105017887) related to the proposed waveguide structure in this work.

\section*{Data availability Statement}
The data that support the findings of this study are available from the corresponding authors upon reasonable request.

\section*{Acknowledgements} 
This work was supported by National Natural Science Foundation of China (62575017), Postdoctoral Fellowship Program (GZB20230009), and Fundamental Research Funds for the Central Universities (RCPT-6120200160). The authors would like to thank the support from Center of Intelligent Sensing and Precise Testing Technology in Optics and Photonics at Beijing Institute of Technology.

\clearpage
%
\bibliographystyle{MSP}
\bibliography{ref}





\end{document}